# Revisiting the field normalization approaches and practices

Xinyue Lu[1,2], Li Li[1], and Zhesi Shen[1,2]

[1] Natiaonal Science Library, Chinese Academy of Sciences, 100190 Beijing, China
[2] Department of Information Resources Management, School of Economics and Management, University of Chinese Academy of Sciences, Beijing 100190, China

*Email addresses:* luxinyue@mail.las.ac.cn(X. Lu), lili2020@mail.las.ac.cn (L. Li), shenzhs@mail.las.ac.cn (Z. Shen).

Corresponding author: Zhesi Shen

**Abstract**

Field normalization plays a crucial role in scientometrics to ensure fair comparisons across different disciplines. In this paper, we revisit the effectiveness of several widely used field normalization methods. Our findings indicate that source-side normalization (as employed in SNIP) does not fully eliminate citation bias across different fields and the imbalanced publication growth rates across fields is a key factor for this phenomenon. Recently, a combination of logarithmic transformation and mean-based normalization, expressed as ln(c+1)/mu, has gained popularity. However, our analysis shows that this approach does not yield satisfactory results. Instead, we find that combining logarithmic transformation (ln(c+1)) with z-score normalization provides a better alternative. Furthermore, our study suggests that the better performance is achieved when combining both source-side and target-side field normalization methods.

## 1. Introduction

Citation and its derivative indicators are commonly used to reflect impact and are among the most important quantitative metrics in scientific evaluation (Garfield, 2006). However, differences in citation potential among fields result in field biases in citation-based indicators (Leydesdorff & Bornmann, 2011). The development and improvement of metrics which support cross-field comparison become a crucial issue in scientometrics.

Citation field normalization encompasses two important problems: how to treat the field difference and how to conduct the normalization. As for the first problem, there are two main streams of research aimed at addressing field bias: source-side normalization and target-side normalization.

Theoretical basis of source-normalized methods is that the varying citation density across fields is due to differences in the length of references (Mingers & Yang, 2017; Zitt & Small, 2008). In 2008, Zitt and Small proposed to normalize the raw citation by considering the reference length of citing source (1/r) (Zitt & Small, 2008). Later the concept of active reference (1/a) is introduced in 2011 (Leydesdorff & Bornmann, 2011) and journal's activity factor in 2012 (Waltman et al., 2013; Waltman & van Eck, 2013b), to account for the different accumulation rates of citations across different fields. The prerequisites for source-normalized methods to function fully are overly idealized and cannot be achieved in practice. Waltman (Waltman & van Eck, 2013a) conducted a systematic large-scale empirical comparison among three source-normalized methods, but the evaluation framework he used does not support statistical tests. Meanwhile, how topic growth relates to citation counts and impacts citing-side normalization (Leydesdorff & Opthof, 2010; Waltman et al., 2013; Waltman & van Eck, 2013b) is not intuitive and still not well understood (Sjögårde & Didegah, 2022). This gap highlights the need for further research to explore how topic growth dynamics influence citation patterns and normalization practices.

The primary idea behind target-normalized methods is to calculate a relative citation performance given a comparable set for each publication or journal, which is commonly based on a field classification system (Leydesdorff & Bornmann, 2011). Therefore, the first issue of this normalized approach lies in the selection of field classification system (Bornmann, 2020). Recent studies have shown that a paper-level classification system performs better than journal-level classification system in reducing the citation bias (Ruiz-Castillo & Waltman, 2015; Shu et al., 2019; Strotmann & Zhao, 2010; Tong et al., 2023).

Once the classification system is determined, the second issue is selecting the normalization approaches which typically receives relatively less attention but is crucial. Currently, mean-based normalization ($c/\mu$) (Abramo et al., 2012a, 2012b; Radicchi et al., 2008) and z-score transformation ($c\text{-}\mu/\sigma$) are the widely used practices because they are intuitive and simple. Recently, the log transformation of citation is introduced to overcome the skewness of citation distribution (Brzezinski, 2015; Eom & Fortunato, 2011; Lundberg, 2007; Shen et al., 2018; Stringer et al., 2008). Furthermore, the normalization is applied to the transformed citation, especially the z-score normalization approach (Lundberg, 2007). A more detailed discussion of normalization approaches, including their applications and limitations, can be found in the review in 2016 (Waltman, 2016). Here leads to our second question, does this combination result better normalization performance (log-transformation + Z-score), or which type combination performs better?

In this paper, we want to answer the following four key questions:

(1) Can the source-normalized methods entirely eliminate citation bias among fields?

(2) For target-side normalization, among c/μ, c-μ/σ, ln(c)/μ, ln(c)-μ/σ, which approach has better performance?

(3) Will the combination of source-side and target-side normalization achieve better performance?

(4) How do different citation normalization approaches affect journal evaluation？

## 2. Data and Methods

### 2.1 Publication data and citation data

We collect articles and reviews indexed in the Web of Science (WoS) between 2020 and 2021 and their citations received in 2022. To ensure consistency in the data coverage, we focus exclusively on articles and reviews indexed in the Science Citation Index Expanded (SCIE) and Social Sciences Citation Index (SSCI) categories. Finally, the dataset for 2020 and 2021 comprises a total of 4,520,494 papers, while the dataset for 2022 includes 2,221,501 papers. Additionally, the citation relationships in 2022 contain 118,294,005 citations, with 16,329,497 of these citations referencing core papers published in 2020 and 2021.

### 2.2 Field Classification systems

In this study, we leverage two distinct field classification systems to categorize the collected publications, ensuring a more robust and unbiased approach to field normalization and evaluation. Specifically, we align the papers in our dataset with both the CWTS paper-level classification system, which provides hierarchical classification across three granularity levels—micro-level, meso-level, and macro-level topics (Waltman & van Eck, 2012)—and the SciSciNet subfield classification system, which is derived from the MAG (Microsoft Academic Graph) dataset and consists of 292 specific subfields (Lin et al., 2023). This dual-classification strategy addresses the potential issue of bias that may arise when using a single classification system for both normalization and evaluation, thus avoiding the “athlete and referee” situation, where the same classification system influences both the standardization and assessment processes. Unlike CWTS, the SciSciNet system can assign a publication to multiple subfields. For clarity and consistency, in all main-text analyses we use the first assigned subfield label as the publication’s primary field. To ensure robustness, we also repeated all analyses by assigning each publication to all of its associated subfields (shown in Supplematry Materials). The results, reported in the Supplementary Materials, remain consistent with those presented in the main text, confirming that our findings are not sensitive to the choice of subfield selection strategy.

The full data processing pipeline is illustrated in Figure 1. First, publications were matched with the CWTS classification system using their UT (unique article identifier), achieving a 90.9% match rate. Second, publications were matched with the SciSciNet classification system via DOI, achieving a 97.9% match rate. For the remaining unmatched publications, we employed SPECTER-based text embeddings (Cohan et al., 2020) combined with the k-nearest neighbor(KNN) algorithm to assign them to the most relevant topics, which ensures that nearly all collected publications are assigned with a topic in both field classification systems.

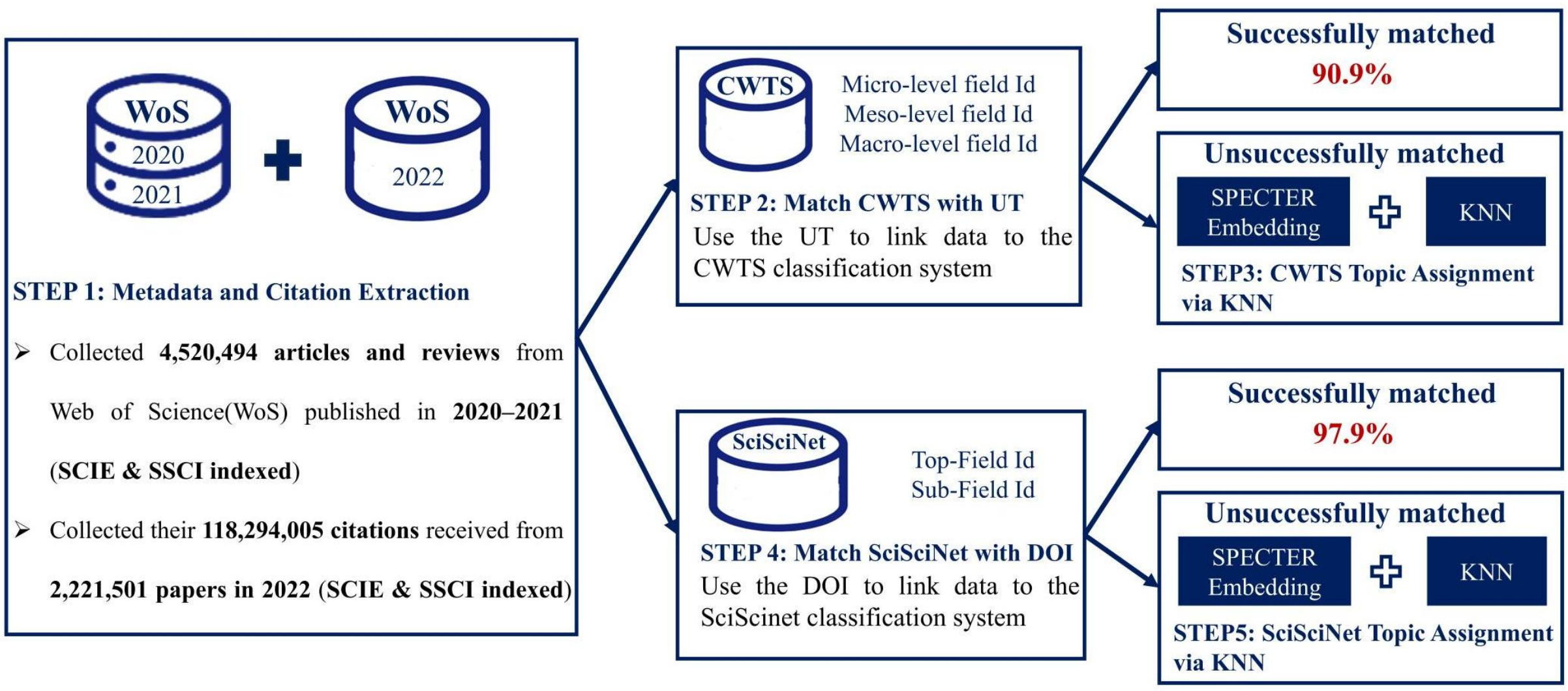

**Figure 1.** Data processing pipeline for publication collection, matching and classification.

## 2.3 Citation Indicators

Building on the normalization approaches discussed earlier, we describe and define the key citation metrics that will be used in our analysis. The normalization methods are put into three categories: source-side metrics, target-side metrics, and dual-side metrics. Each category offers different approaches to adjust for field-specific biases.

*Unnormalized metric*

Citation count, *c*. The citation count refers to the citations received by paper *i* in a given year.

*Source-side normalized metrics*

① First source-side normalized citation count, $sc^{(1)}$. The $sc_i^{(1)}$ value of paper *i* is calculated as:

$$sc_i^{(1)} = \sum_{i=1}^{c_i} (\frac{1}{r_i}) ,$$

where $r_i$ is the length of reference list in the paper from $i^{th}$ citation. $sc^{(1)}$ would suppress citation bias among fields from source theoretically (Waltman et al., 2013).

② Second source-side normalized citation count, $sc^{(2)}$. The value $sc_{\mathrm{i}}^{(2)}$ of paper *i* is calculated as:

$$sc_i^{(2)} = \sum_{i=1}^{c_i} (\frac{1}{a_i}) ,$$

where $\boldsymbol{a_i}$ is the number of active references in the paper from which $\boldsymbol{i^{th}}$ citation generates. Active reference is defined as papers in Web of Science, falling into the time window of analysis year (Waltman & van Eck, 2013b; Zitt & Small, 2008). For example, the active reference length for the 2-year time window of publications in 2022 refers to the number of references publishing between 2020 and 2021.

③ Third source-side normalized citation count, $sc^{(3)}$. The $sc_i^{(3)}$ value of paper *i* is calculated as:

$$sc_i^{(3)} = \sum_{i=1}^{c_i} (\frac{1}{a_i \times p_i}),$$

where the definition of $a_i$ is the same as $sc^{(2)}$ and $p_i$ is the proportion of publications which contains at least one active reference among all publications in journal of $i^{th}$ citing publication (Waltman et al., 2013).

For the above four metrics, we also calculate their logarithmic form: $\ln(c_i + 1)$, $\ln(sc_i^{(1)} + 1)$, $\ln(sc_i^{(2)} + 1)$ and $\ln(sc_i^{(3)} + 1)$, and respectively defined them as $c^{\ln}$, $sc^{(1)\ln}$, $sc^{(2)\ln}$ and $sc^{(3)\ln}$.

*Target-side normalized metrics*

For target-side normalized metrics, we consider two normalize approaches: relative ratio and z-score for *c* and ln(*c*+1).

① Relative ratio, $r^f$, we define it as

$$r_i^f = \frac{m_i}{\mu^f},$$

where $m_i$ refers to metric value of paper *i* and $\mu^f$ is average metric value of papers which belongs to the same field, document type and publication year with paper *i*.

② z-score, $z^f$ .We define it as

$$z_i^f = \frac{m_i - \mu^f}{\sigma^f},$$

where $\mu_i^f$ is average metric value of papers which belongs to the same field with paper *i* and $\sigma^f$ is the standard deviation of metric value in field *f*.

*Dual-side normalized metrics*

By combining source-side and target-side normalization approaches, we have the dual-side normalized metrics as shown in Table 1.

Table 1 shows the total indicators we investigated in this work. We combine Citation count, $c$ and three source-side metrics with two different normalized approaches, resulting in 24 metrics (Table 1). The structure of Table 1 represents the categories to which the normalization methods used for each metric belongs (non-normalized, source-normalized, target-normalized or both).

**Table 1.** The combination of citation-based metrics and normalized approaches.

| - | | *None* | | *Source side normalization* | | | | | |
|---|---|---|---|---|---|---|---|---|---|
| | | *Original* | *Log* | *Original* | | | *Log* | | |
| None | - | $c$ | $c^{\ln}$ | $sc^{(1)}$ | $sc^{(2)}$ | $sc^{(3)}$ | $sc^{(1)\ln}$ | $sc^{(2)\ln}$ | $sc^{(3)\ln}$ |
| Target side Normalization | Ratio | $R(c)$ | $R(c^{\ln})$ | $R(sc^{(1)})$ | $R(sc^{(2)})$ | $R(sc^{(3)})$ | $R(sc^{(1)\ln})$ | $R(sc^{(2)\ln})$ | $R(sc^{(3)\ln})$ |
| | Z-score | $Z(c)$ | $Z(c^{\ln})$ | $Z(sc^{(1)})$ | $Z(sc^{(2)})$ | $Z(c^{\ln})$ | $Z(sc^{(1)\ln})$ | $Z(sc^{(2)\ln})$ | $Z(sc^{(3)\ln})$ |

### 2.4 Normalization performance evaluation

To assess the effectiveness of each metric in reducing citation field bias, we adopt three complementary approaches: (1) a correlation check for intuitive inspection, (2) Mahalanobis distance for statistical evaluation, (3) Kullback–Leibler (KL) divergence for an information-theoretic perspective.

*Evaluating bias among fields: correlation check*

The first qualitative method is based on a simple intuition: the average of field normalized citation count of each topic shouldn't show a strong positive correlation with the average of raw citation count. we therefore conduct scatter plots for each metric using field normalization methods against citation count to observe the relationship between them.

*Evaluating bias among fields: Mahalanobis distance*

The second quantitative method is grounded in the following assumption: if the rankings derived from a given metric are not biased across scientific fields, then the proportion of publications from each field within the top z% of ranked publications should match the proportion of that field in the entire dataset (Dunaiski et al., 2019; Vaccario et al., 2017). In other words, publications from each field should be evenly distributed across every ranking interval. To quantitatively assess this deviation, we adopt the evaluation standard $d_{\mathrm{M}}$ proposed by Vaccario (Vaccario et al., 2017).

Specifically, we compute the distributional inequality between the observed field representation in the top z% and the expected distribution under field-neutral conditions. The greater this discrepancy, the poorer the effect of field normalization.

For a given metric $m$, the expected number of papers from subfield $i$ in the top $z\%$ under perfect field normalization is $\mu_i^{(m)} = (z/100) \cdot K_i$, where $K_i$ is the total paper numbers in subfield $i$. The observed count $k_i^{(m)}$ represents the actual representation of subfield $i$ in the top $z\%$. Then we can quantify the overall field bias using the Mahalanobis distance ($d_{\mathrm{M}}$):

$$d_M^{(m)} = \sum_{i=1}^{F} \frac{(k_i^{(m)} - \mu_i^{(m)})^2}{\sigma_i^2} \cdot (1 - \frac{K_i}{N}),$$

where $\sigma_i^2 = \gamma \cdot K_i \cdot (N - K_i)$ is the expected variance and $N$ is the total papers in the dataset. The finite-population correction factor $\gamma = \frac{n \cdot (N-n)}{N^2 \cdot (N-1)}$ accounts for the reduced variance in sampling without replacement, ensuring cross-sample comparability of bias measurements. The term $(1 - \frac{K_i}{N})$ dampens the disproportionate influence of dominant subfields on the aggregate bias metric, preventing overestimation from majority fields.

It is worth noting that we utilized the micro-level topics from CWTS to standardize various metrics on the target side, while meso-level topics was employed to compute $d_{\mathrm{M}}$ to evaluate the effectiveness of the standardization. Additionally, we also used the subfield classification system from SciSciNet to recalculate $d_{\mathrm{M}}$ as a robustness check.

*Evaluating bias among fields: Kullback–Leibler (KL) divergence*

As a complementary perspective, we also incorporate the Kullback–Leibler (KL) divergence, which provides an information-theoretic measure of the discrepancy between observed and expected field distributions (Gates et al., 2024; Ke et al., 2023).

For each metric $m$:

$$P_i^{(m)} = \frac{k_i^{(m)}}{n},$$

$$Q_i = \frac{K_i}{N},$$

where $P_i^{(m)}$ denotes the observed proportion of subfield $i$ in the top z% and $Q_i$ is its expected proportion in the full dataset. The KL divergence is then defined as:

$$D_{KL}^{(m)} = \sum_{i=1}^{F} P_i^{(m)} \cdot log \frac{P_i^{(m)}}{Q_i}.$$

This measure quantifies the information loss when the expected distribution is used to approximate the observed distribution, with larger values indicating poorer field normalization.

In practice, the two approaches are grounded in similar principles, but the Kullback–Leibler (KL) divergence offers a complementary perspective and serves as a robustness check for the conclusions.

# 3. Results

### 3.1 RQ1: Can the source-normalized methods entirely eliminate citation bias among fields?

To answer this question, we calculate the average normalized citation ($sc^{(1)}$, $sc^{(2)}$ and $sc^{(3)}$) of each micro topic against the average of raw citation count $c$ as shown in Fig.2. A high pearson and spearman correlation means that the citation advantages of topics is not eliminated after the field normalization approaches. We can see in Fig.2, among the source-normalized indicators, $sc^{(3)}$ has the smallest correlation coefficient suggesting it is more effective in field normalization compared with the other two indicators. However, all three source-normalized methods ($sc^{(1)}$, $sc^{(2)}$ and $sc^{(3)}$) still show a strong positive correlation with citation count ($c$), suggesting that none of the three source-normalized indicators fully eliminate the field biases.

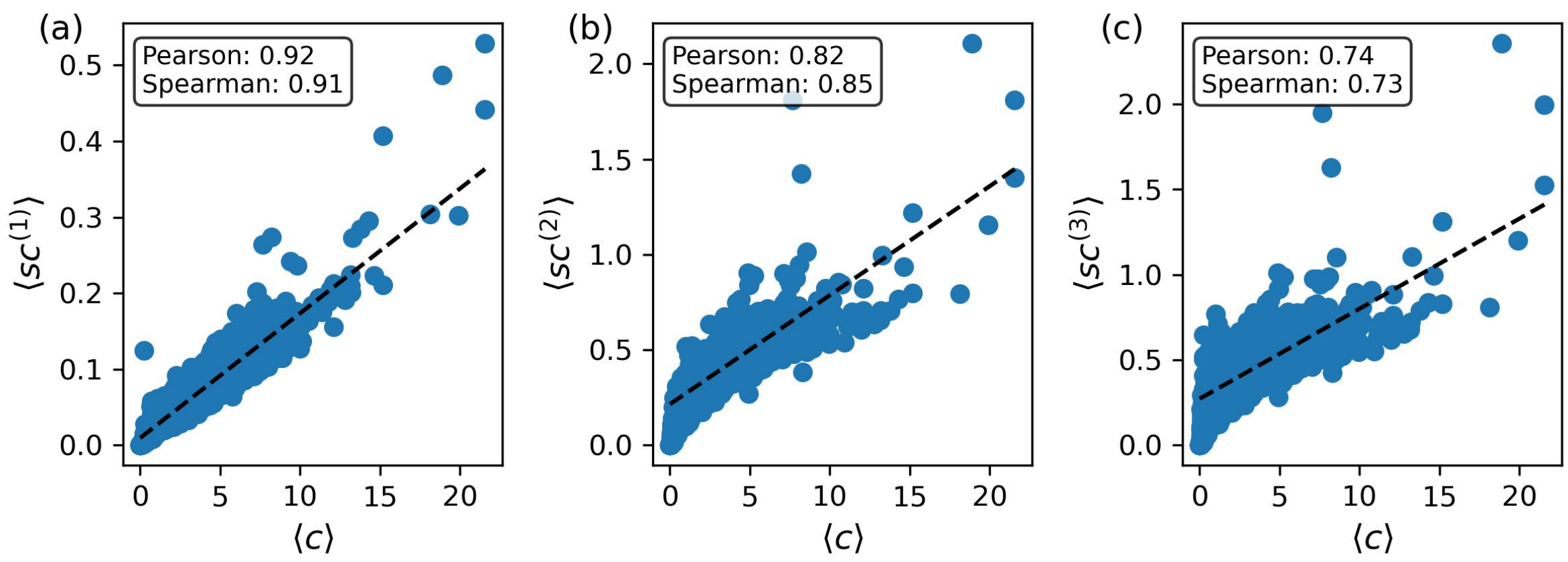

**Figure 2.** Correlation between the average of citation count and source-normalized metrics of each topic.

We further validate the conclusion through a quantitative evaluation method based on $d_M$ and KL divergence. The smaller the $d_M$ or KL divergence, the better the normalization effect of metric among fields. As shown in Figure 3, in both classification schemes and for both evaluation methods, the results are consistent: $sc^{(3)}$ achieves the best performance across all percentiles, followed by $sc^{(2)}$, and then $sc^{(1)}$. All three methods perform better than the benchmark $c$, demonstrating a certain degree of effectiveness of the source-normalized methods in reducing field bias regardless of the classification scheme or evaluation metric used.

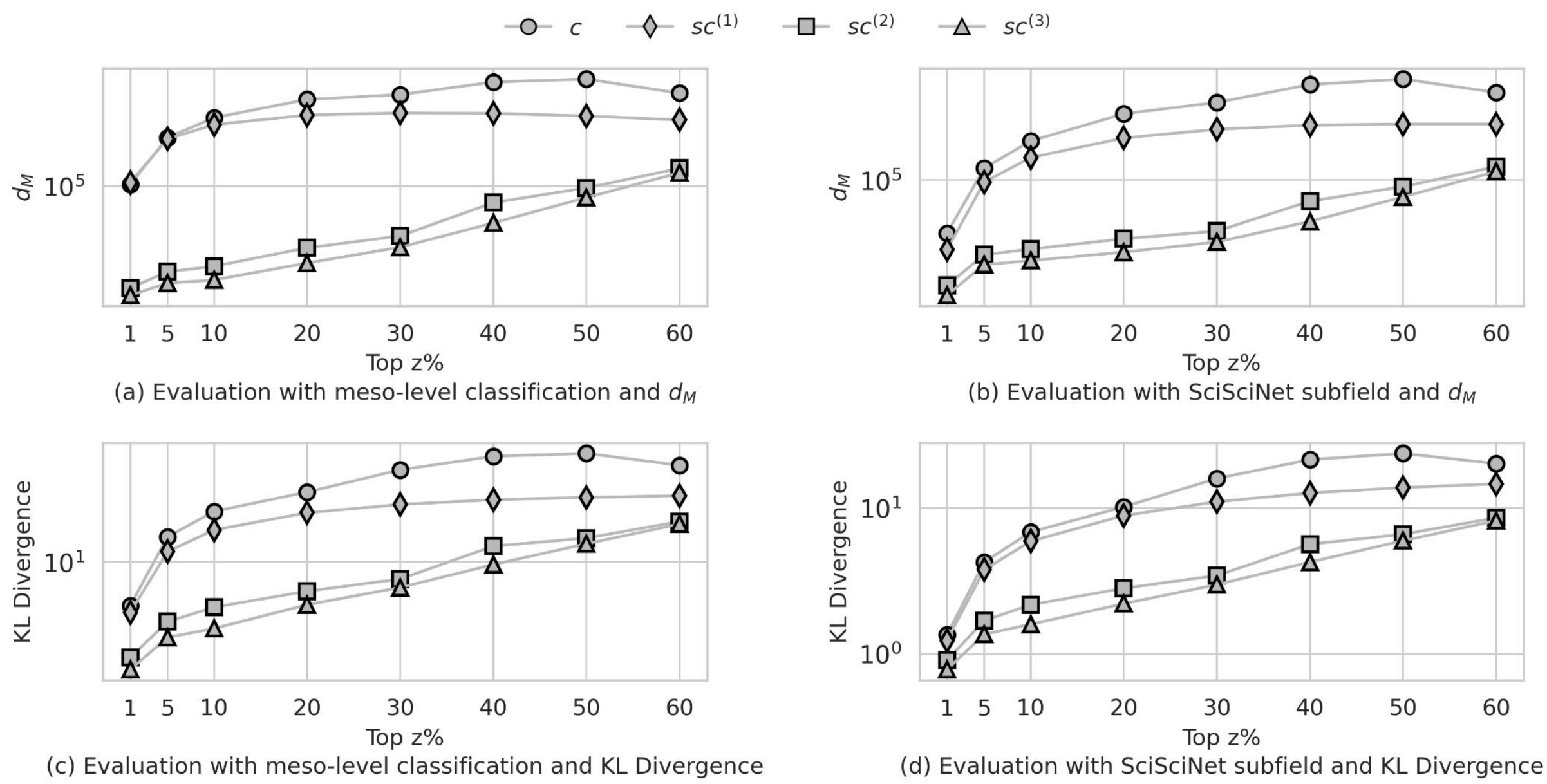

**Figure 3.** Field bias of source-normalized metrics.

We further investigate why source-side normalization methods fail to fully eliminate field bias in the normalization process. According to the calculation of SNIP (Waltman et al., 2013), there are three key assumptions for ensuring the effectiveness of source-side normalization:

(1) the same number of papers are published annually within each field, i.e., $M^f_{2020} = M^f_{2021} = M^f_{2022}$;

(2) there is no citation overlap between journals from different fields;

(3) each journal has at least one paper with an active reference.

If these three assumptions hold, the mean value ($\mu$) of $sc^{(3)}$ for each field can be calculated as shown in the following formula (for details, see paper (Waltman et al., 2013)):

$$\mu = \frac{2(M_{2020} + M_{2021})}{M_{2022}} = 1.$$

The first two assumptions are difficult to achieve in practice and may help explain why these metrics fail to perform as expected. To further explore this issue, we test the validity of the first two assumptions.

The core of assumption 1 is that $M_{2022} = \frac{1}{2}(M_{2020} + M_{2020})$, implying that the number of papers published in a given field in 2022 should be equal to half of the total number of papers published in 2020 and 2021. However, in reality, the number of papers published in each field fluctuates every year, with varying degrees of change across different fields. This variation results in a mean value for $sc^{(3)}$, that deviates from 1. To quantify this variation, we define the *growth rate* of a field as $\frac{(M_{2020} + M_{2021})}{M_{2022}}$. A higher growth rate corresponds to a larger μ value for that field.

Assumption 2 posits that there is no citation overlap between journals from different fields. The core of source-normalized methods is the adjustment of citation counts by dividing them by the citation density of the corresponding field. If a paper is cited by journals from other fields, the citation density is either overestimated or underestimated, leading to normalization failure. To test this assumption, we define the *citation density* of a field.

The citation density of a field *f*, denoted as $D_f$ , is defined as the total number of active references generated by all papers within the field. For a given paper $i$, its actual citation density $AD_i$ is calculated as the weighted average of the citation densities of the fields that cite it:

$$AD_i = \sum_{k \in CitingFields_i} w_{i,k} \cdot D_k,$$

where $w_{i,k}$ is the proportion of citations received by paper $i$ from field $k$ and $D_k$ is the citation density of field $k$. The expected citation density of paper $i$ is the citation density of its own field $f$, denoted as $D_f$. Based on these, the density ratio for paper $i$ is defined as:

$$DR_i = \frac{AD_i}{D_f}.$$

A ratio greater than 1 indicates that a paper's citation density is overestimated, which may lead to an underestimation of its true impact. Conversely, a ratio less than 1 suggests the citation density is underestimated, potentially overestimating the paper's impact.

Figure 4(a) shows the mean value of $sc^{(3)}$ for each meso-field in the CWTS classification against the growth rate. The colour of each data point represents the citation density ratio. We observe a clear positive correlation between the growth rate and the mean $sc^{(3)}$, and we find that when the citation density ratio is higher, the mean $sc^{(3)}$ tends to be smaller. Figure 4(b) demonstrates a positive correlation between the mean value of $sc^{(3)}$ and the growth rate, while showing a negative correlation with the citation density ratio. Further residual analysis reveals that the growth rate explains 63.7% of the variance in the mean value of $sc^{(3)}$, suggesting that it is a primary factor contributing to the failure of field normalization.

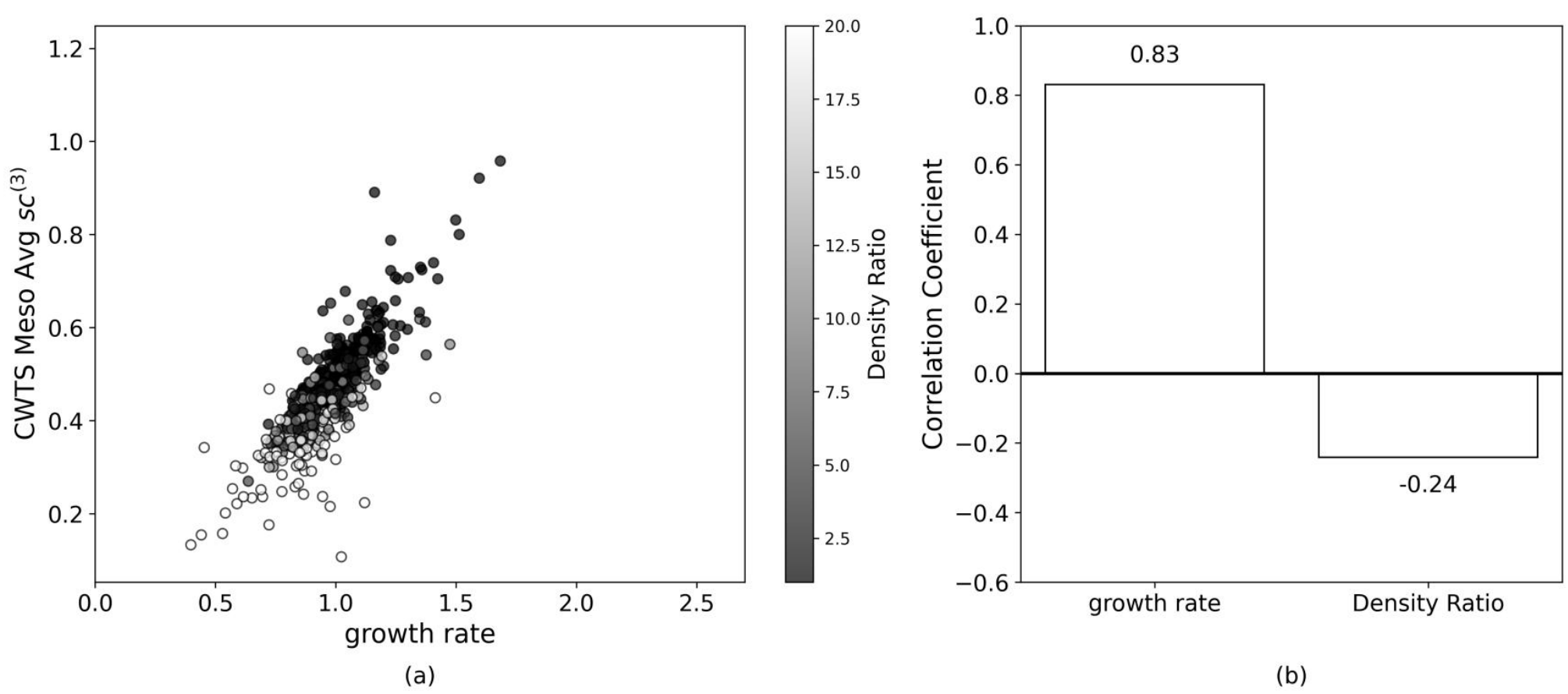

**Figure 4. Factors affecting the effectiveness of source-side normalization. (**a) Correlation between growth rate and average c for CWTS meso-level fields. (b) Strength of Correlation between Growth Rate/Density Ratio and Average $sc^{(3)}$ by CWTS meso-level fields.

### 3.2 RQ2: Among c/μ, c-μ/σ, ln(c)/μ, ln(c)-μ/σ, which approach has better performance?

To explore the effectiveness of different target-normalization methods, we considered the ratio-normalized original form $R(c)$, z-score-normalized original form $Z(c)$, ratio-normalized logarithmic form $R(c^{\ln})$, and z-score-normalized logarithmic form $Z(c^{\ln})$ and evaluate their normalization performance under meso-level classification and SciSciNet subfield as shown in Figure 5. The results indicate that no single method dominates across all scenarios. Nonetheless, two approaches perform relatively better on average: ratio normalization of raw citation counts $R(c)$ and z-score normalization of logarithmic counts $Z(c^{\ln})$.

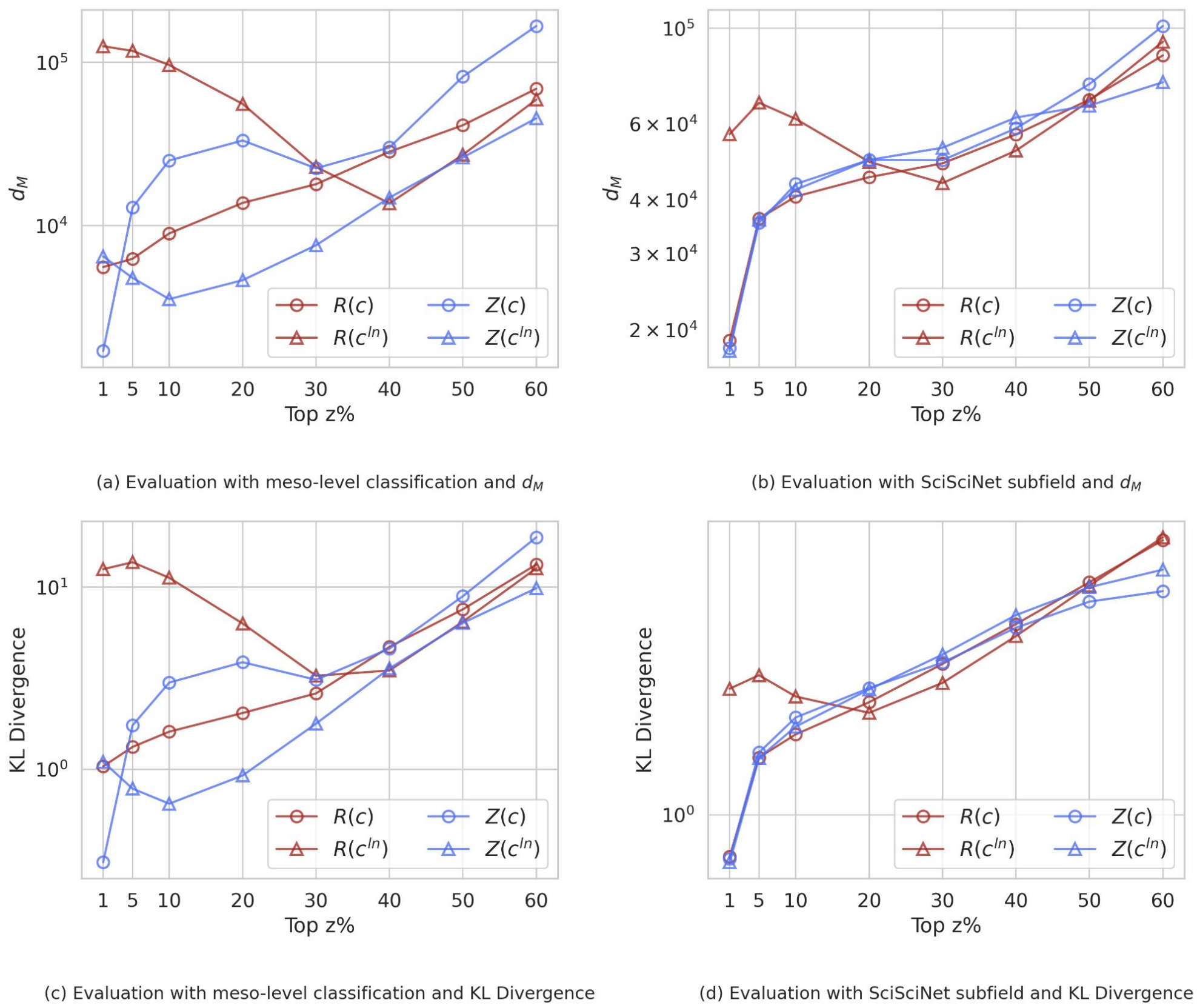

(a) Evaluation with meso-level classification and $d_M$

(b) Evaluation with SciSciNet subfield and $d_M$

(c) Evaluation with meso-level classification and KL Divergence

(d) Evaluation with SciSciNet subfield and KL Divergence

**Figure 5.** Field normalization performance of different target-side approaches.

Strictly proving the effectiveness of these normalization metrics is challenging, but we can provide an intuitive explanation. Citation distributions are often approximated as log-normal distributions (Shen et al., 2023; Stringer et al., 2008). Under the logarithmic transformation, there is a natural connection between ln(c) - ln(μ) and ln(c/μ), leading to similar performance for $R(c^{\ln})$ and $Z(c^{\ln})$. Additionally, since the variance across distributions is also considered for $Z(c^{\ln})$, the normalization performance is further improved.

In contrast, applying ratio normalization directly to ln(c), $R(c^{\ln})$ —forces the means of different fields to coincide but unintentionally amplifies the relative variance of low-mean fields. This variance inflation gives these fields an artificial advantage in top-rank publications and reduces the overall normalization quality.

To show how this happens, we use a simple simulation(Figure 6) with three fields differing only in their mean ln(c+1) values. After applying $R(c^{\ln})$, the lowest-mean field becomes substantially more dispersed, because ratio scaling inflates relative variance when the denominator (the mean) is small. In contrast, $Z(c^{\ln})$ aligns both means and variances(Fig.6C). Consequently, the resulting $d_{\mathrm{M}}$ values (Figure 6(d)) show that $R(c^{\ln})$ produces substantially larger deviations than $Z(c^{\ln})$. This simulated data closely mirrors the empirical results in Figure 5, providing a clear visual explanation for why ratio normalization on log-transformed citations is theoretically unsound and practically unreliable.

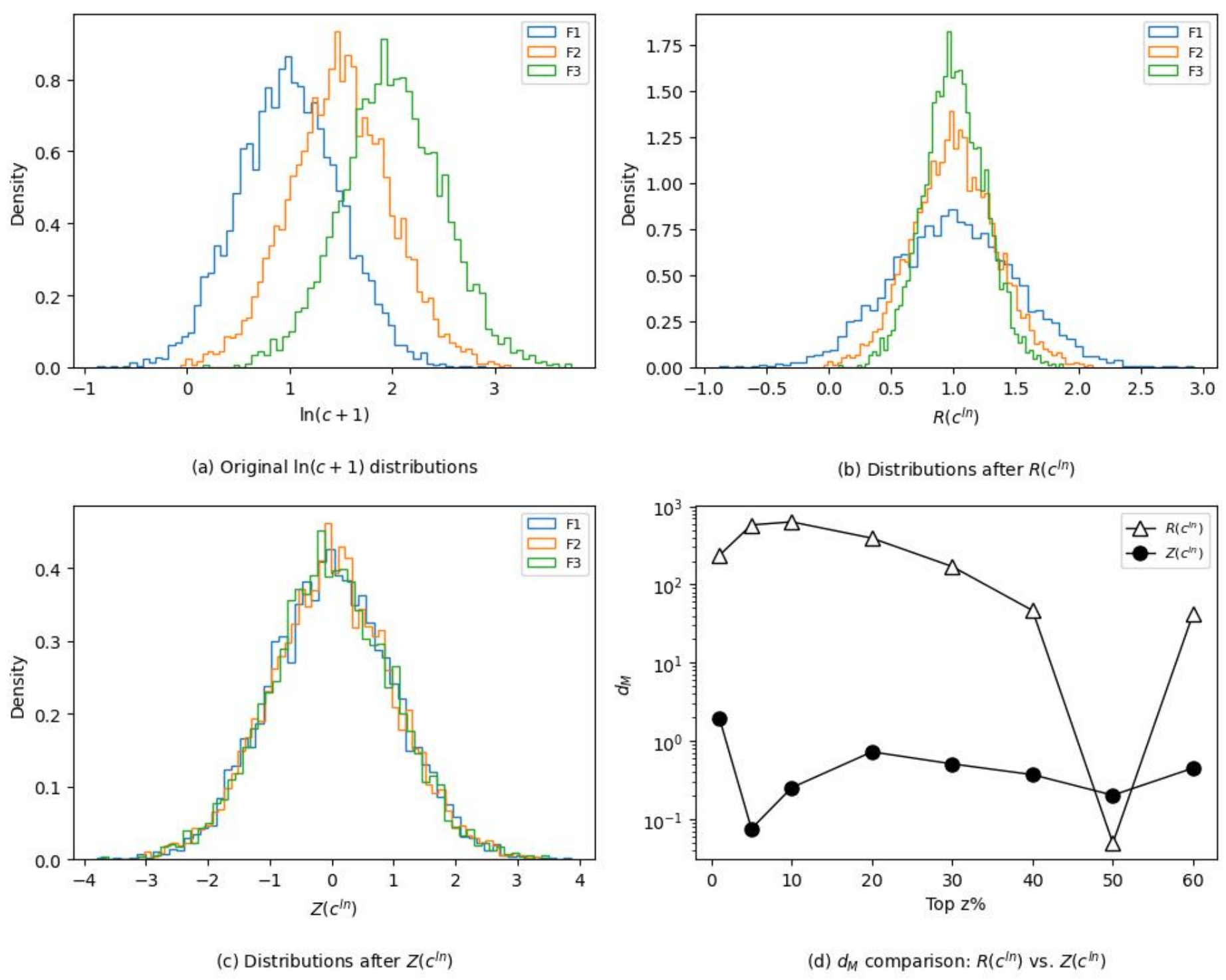

**Figure 6.** Simulation-based comparison of log-ratio and log-z-score normalization.

### 3.3 RQ3: Will the combination of source-side and target-side normalization achieve better performance?

To address the question of whether combining source-side and target-side normalization can yield better performance than using source-side normalization alone, we leveraged the conclusions from *RQ1* and *RQ2*. *RQ1* demonstrated that $sc^{(3)}$ is the most effective source-normalized metric, and *RQ2* showed that applying ratio normalization to the original citation counts or using log-transformed z-score normalization generally yields better results. Building on these findings, we combined with the ratio normalization and log-transformed z-score methods to create two new indicators: $R(sc^{(3)})$ (ratio-normalized) and $Z(sc^{(3)\ln})$ (log-transformed z-score-normalized). These newly constructed indicators were then compared against existing indicators, including $c$, $R(c)$, and $Z(c^{\ln})$, to evaluate their relative effectiveness in normalizing citation data and reflecting a paper's impact within its field.

As illustrated in Figure 7, two robust patterns emerge. First, across all evaluation settings, the combined indicators consistently perform better than source-only and target-only approaches. Specifically, the triangular markers (combined methods) always lie below the circles (source-only) and stars (target-only) across all percentile thresholds. Second, among all methods, the combination of z-score normalization with the logarithmic form of $sc^{(3)}$ consistently provides the strongest normalization effect. The second-best method, however, differs by classification system: under meso-level evaluation, $Z(c^{\ln})$ performs better, whereas under the SciSciNet subfields, $R(sc^{(3)})$ ranks next.

Overall, these findings highlight that integrating source-side and target-side normalization substantially improves performance, with the z-score–logarithmic combination being the best for addressing field citation biases.

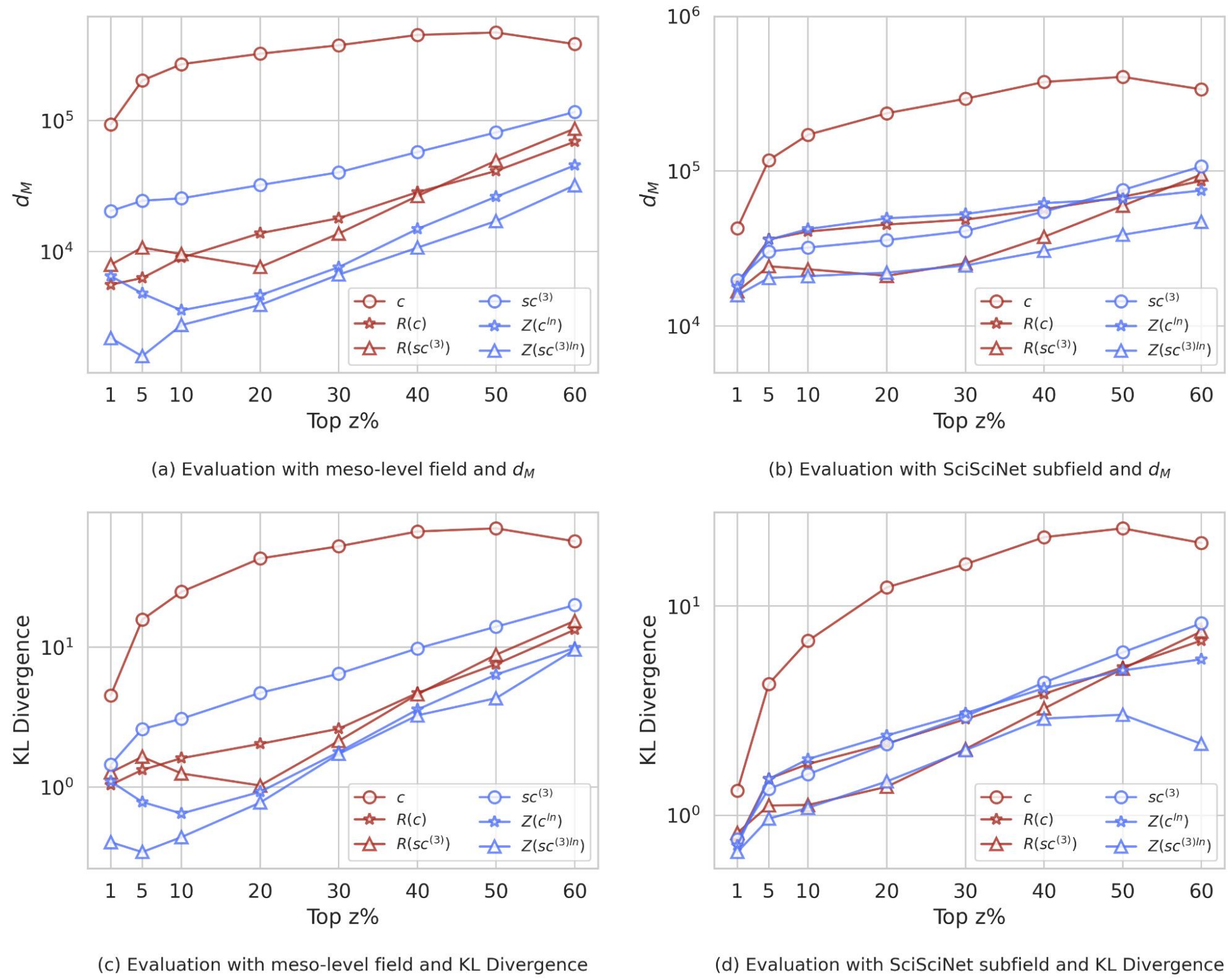

(a) Evaluation with meso-level field and $d_M$

(b) Evaluation with SciSciNet subfield and $d_M$

(c) Evaluation with meso-level field and KL Divergence

(d) Evaluation with SciSciNet subfield and KL Divergence

**Figure 7.** Field normalization performance for combing source and target-side approaches.

To better understand the source of these improvements, we examined whether the observed gains from the combined method were uniformly distributed across fields or concentrated in specific domains. We compared the benchmark shares of the five macro-level fields under the raw citation, target normalization ($Z(c^{\ln})$) and combined normalization ($Z(sc^{(3)\ln})$) (Figure 8 a). Results indicate that raw citation counts substantially distort field representation in the top 5%, while both normalization strategies mitigate this bias. However, target-only normalization tends to overcorrect the bias, whereas the combined method achieves a more balanced adjustment across fields.

At the meso-level, we compared the $d_{\mathrm{M}}$ under $Z(c^{\ln})$ and $Z(sc^{(3)\ln})$ (Figure 8b). Among 822 CWTS meso-level fields, 562 (68.37%) achieved lower $d_{\mathrm{M}}$ with the combined approach, indicating closer alignment with expected distributions. Figure 7b highlights the ten meso-level fields with the largest reductions in $d_{\mathrm{M}}$. These ten disciplines (10/822, 1.22%) account for 20.77% of the total reduction in $d_{\mathrm{M}}$, indicating that the observed improvements are focused on a small set of fields, rather than evenly distributed.

For these ten meso-fields, Figure 8c compares the deviations of these fields' representation in the top 5% of papers under citation counts, $Z(c^{\ln})$ and $Z(sc^{(3)\ln})$ from their benchmark shares. Most fields belong to the mathematical sciences (algebra, discrete mathematics, mathematical analysis) or the humanities and social sciences (constitutional law, linguistics, cross-national migration, Asian philosophy, history of economic thought). These fields typically have low citation density. A notable exception is the life-science domain on non-coding RNA research, which has grown rapidly during the publication and citation period covered by this work (Lou et al., 2024).

The observed patterns likely stem from multiple factors. Predefined target field labels may inadequately capture disciplinary boundaries. This sometimes leads to overcorrection. The source-based approach offers relative advantages in domains with strong within-field citations by leveraging references to construct more accurate comparison sets. In addition, data-specific characteristics, such as within-field citation distributions, may also contribute. These results indicate that the benefits of combining target- and source-based normalization are unevenly distributed across disciplines, suggesting directions for further refinement of field-normalization methods.

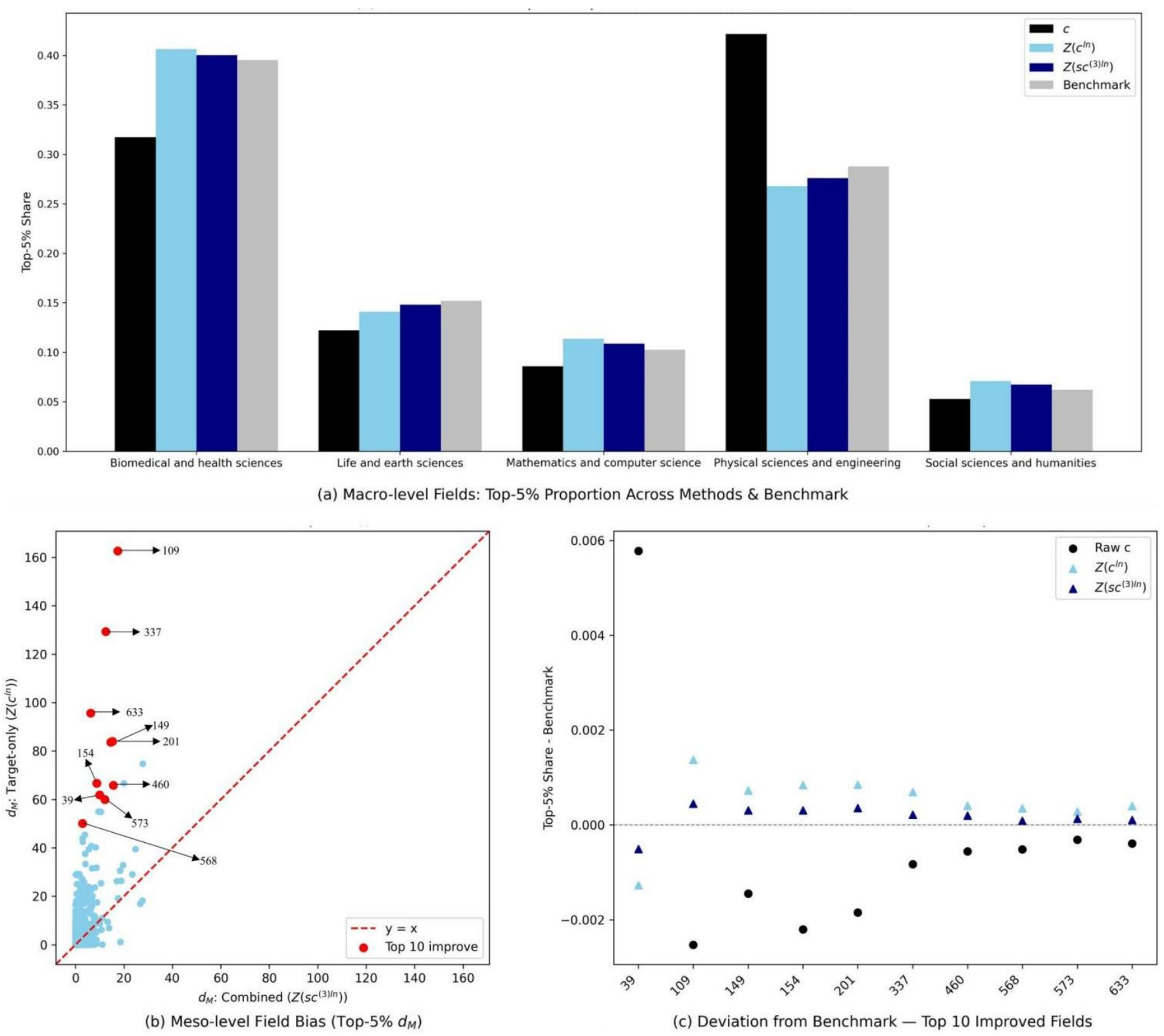

**Figure 8. Field-level analysis of normalization performance.** (a) Macro-level field shares in the top-5% under citation counts, target normalization, and combined normalization. (b) Meso-level field

deviations from the benchmark, with red points marking the ten fields showing the largest reductions in $d_{\mathrm{M}}$ . (c) Detailed deviations for these ten meso-fields: non-coding RNA networks and disease mechanisms (#39), migration and socio-cultural dimensions (#109), discrete mathematics and theoretical computer science (#149), education studies (#154), linguistics and intercultural communication (#201), history of economic thought (#460), mathematical analysis (#568), Asian philosophy and religion (#573), U.S. constitutional law (#337), and algebra (#633).

Note: CWTS provides only descriptive labels for meso-fields; the textual names reported here were generated using a large language model based on these descriptions (see Supplementary Materials).

### 3.4 RQ4: How do different citation normalization approaches affect journal evaluation?

To further examine how different normalization approaches shape journal evaluation outcomes, we extended the analysis from individual papers to journals. The journal-level comparison allows us to identify where different normalization schemes diverge and to what extent they alter the relative standing of journals across different disciplines. To assess the consistency among different normalization indicators, we first ranked all journals based on the scores derived from each normalized citation indicator tested in our study. And we then divided these ranked journal into four distinct citation performance zones: Zone I(top 5%), Zone II(5-20%), Zone III(20-50%), and Zone IV(50-100%). Finally, we calculate the fraction of journals shared(overlap) betweent the corresponding sets across all tested indicators within each of these four zones using a Top-k coverage approach.

The intersection analysis, illustrated in Fig.9 reveals a distinct pattern across the four zones. At the top tier (Zone I), indicators display strong convergence, with overlap ratios between 0.7 and 0.9. This suggests that for the most high-performance journals, the choice of normalization method has a limited influence on their top-tier standing. In the middle tires (Zones II and III), the overlap between raw citation counts and those based on normalized measures drops notebly, typically ranging from 0.5to 0.6,.However, different normalized indicators remained highly consistent with each other (around 0.8-0.87). In the lowest zone, convergence increase again to above 0.8. These findings indicate that normalization has the strongest impact on journal rankings in the middle segments. In contrast, the highest and lowest segments of the ranking are much less affected.

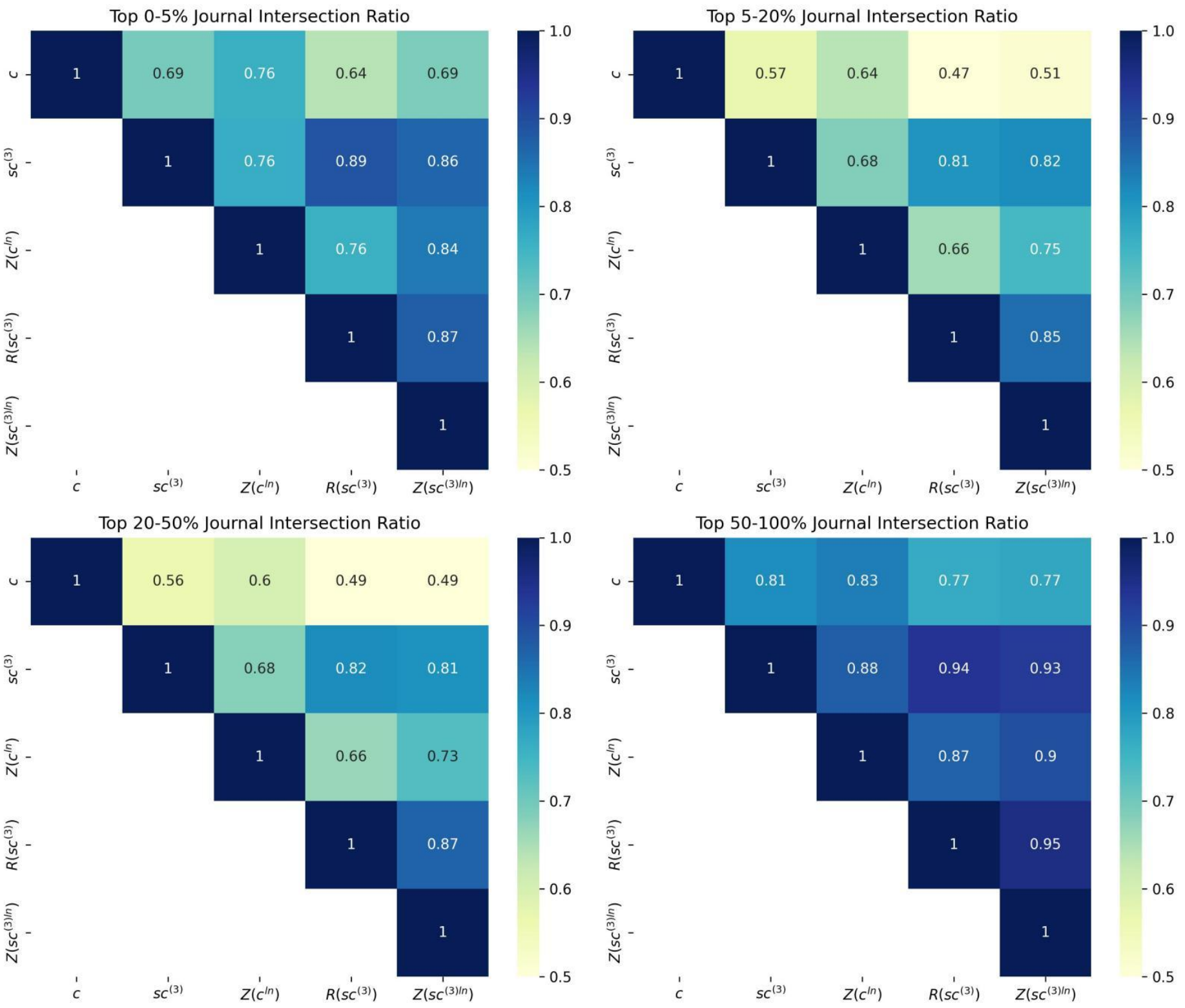

Figure 9 Pairwise intersection ratios of top-ranked journals across normalization indicators (Zones I－IV)

We further explored the disciplinary composition of journals ranked in the top 5% to illustrate the structural consequences of normalization as shown in Fig.10. The inclusion of source-side information slightly reduces the dominance of citation-intensive fields and increases the representation of low-citation domains. The share of Medical & Health Sciences journals decreases from 41.1 % under target-only normalization $Z(c^{\ln})$ to 39.5 % under the combined normalization $Z(sc^{(3)\ln})$, while Humanities & Social Sciences increase from 12.0 % to 13.8 %. The proportions of Natural Sciences, Engineering & Technology, and Multidisciplinary journals remain largely unchanged.

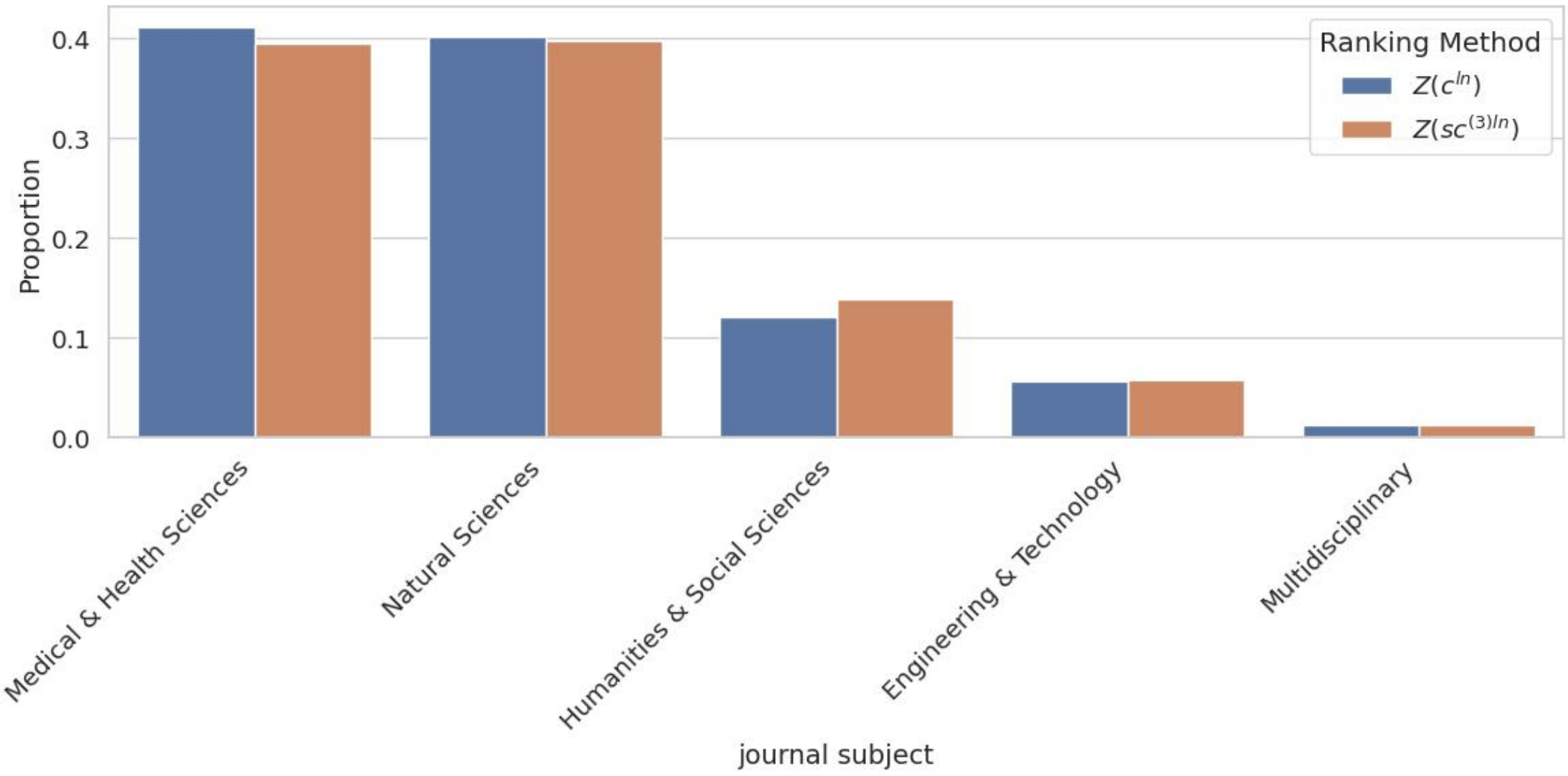

Figure 10 Disciplinary composition of journals in the top 5% under different normalization approaches.

As a case study, we examined journals whose zone assignments changed by at least two tiers between the target-only and combined normalization schemes (Table 2). These large shifts highlight where normalization exerts the most pronounced redistributive effects. Upward movements are concentrated in law and linguistics journals—such as Cambridge Law Journal and Linguistic Inquiry—which rise from Zone IV to Zone II, reflecting adjustments for field-specific citation disadvantages.

Beyond disciplinary redistribution, the inclusion of source-side normalization may also help reducing the effect of possible citation manipulation. Source-side normalization significantly reduces the weight of citations originating from papers with very long reference lists. This diminishes the effectiveness of citation inflation strategies based on excessive referencing. For example, as shown in Table 2, Radiologia Medica, an early-warning journal in 2024 due to suspected citation manipulation[1], presents a pronounced drop after normalization. Overall, these cases illustrate how different normalization approaches reshape journal rankings through selective rebalancing rather than uniform adjustment, mitigating both disciplinary bias and strategic citation behavior.

**Table 2 Journals with zone changes of at least two tiers between target-only and combined normalization**

| **Journal** | Zone | Zone |
|---|---|---|

[1] Early Warning Journal List 2024: https://earlywarning.fenqubiao.com/#/en/early-warning-journal-list-2024

| | under $Z(c^{\ln})$ | under $Z(sc^{(3)\ln})$ |
|---|---|---|
| ANNUAL REVIEW OF LINGUISTICS | 3 | 1 |
| CAMBRIDGE LAW JOURNAL | 4 | 2 |
| CORPUS LINGUISTICS AND LINGUISTIC THEORY | 4 | 2 |
| CYTOJOURNAL | 4 | 2 |
| GEMS & GEMOLOGY | 4 | 2 |
| GEOCHEMICAL TRANSACTIONS | 4 | 2 |
| HAGUE JOURNAL ON THE RULE OF LAW | 4 | 2 |
| HAHR-HISPANIC AMERICAN HISTORICAL REVIEW | 4 | 2 |
| INTERNATIONAL & COMPARATIVE LAW QUARTERLY | 4 | 2 |
| LINGUISTIC INQUIRY | 4 | 2 |
| NAMES-A JOURNAL OF ONOMASTICS | 4 | 2 |
| NATURAL LANGUAGE & LINGUISTIC THEORY | 4 | 2 |
| NATURAL LANGUAGE SEMANTICS | 4 | 2 |
| PHONOLOGY | 4 | 2 |
| SYNTAX-A JOURNAL OF THEORETICAL EXPERIMENTAL A... | 4 | 2 |
| RADIOLOGIA MEDICA | 1 | 3 |

**Conclusion and Discussion**

In this paper, we evaluated various source-normalized methods and found that while they achieve some success in reducing bias across fields, but none of them can fully eliminate it. Our analysis, indicates that imbalanced paper growth rates across fields is a key factor contributing to the limitations of these methods, which not only addresses the puzzle of why these methods are unable to fully eliminate field bias (Sjögårde & Didegah, 2022) but also opens avenues for future research to develop more refined normalization approaches that can better account for such dynamic factors.

We also found that using ratio normalization on original citation counts and log-transformation followed by z-score normalization both yield relatively strong results. However, directly applying ratio normalization after log-transformation is not a theoretically sound method. As a result, some studies that rely on this method should critically re-evaluate their findings.

Furthermore, we found that combining source- and target-normalized methods outperforms using either alone, with log-transformed z-score normalization ( $Z(sc^{(3)\ln})$ ) achieving the best overall performance. However, these combinations still do not fully eliminate field differences. This suggests that while these combinations show promise, further refinement is needed to reduce biases more effectively.

Importantly, improvements from normalization are not uniformly distributed across disciplines. Gains are concentrated in a limited set of fields, particularly in low-citation-density domains (e.g., mathematics, humanities, and social sciences) and occasionally in biomedical areas with atypical citation surges. When applied to journal-level evaluation, the combined normalization also produced visible redistributive effects, reducing the dominance of citation-intensive fields such as medicine while elevating underrepresented areas like linguistics and law. In addition, the inclusion of source-side normalization shows some promise in reducing the effect of citation manipulation. Recent work has proposed the Unique Citing Documents JIF (Uniq-JIF) (Shen et al., 2024), which penalizes journals that accumulate citations from highly overlapping citing sources. Although its formulation differs, the basic idea is similar to source-side normalization—conditioning impact on the nature of the citing sources. Nevertheless, the practical effectiveness of source-side normalization in identifying or constraining manipulative citation behavior still requires systematic investigation.

Despite the comprehensive evaluation, several limitations remain. First, we have not provided a systematic mathematical explanation for the mechanisms underlying different target-normalization methods or for the observed improvements after combining source- and target-based approaches. Second, the reasons why the addition of source-side information produces uneven gains across disciplines—and why normalization in some fields even deteriorates—remain unclear. Third, our analysis was limited to a short citation window. Comparisons using longer citation windows or a dynamic time window were not tested. We believe that the conclusions derived in this paper are highly likely to extend to conditions involving longer time windows, but this requires further testing and evidence.

Future research should address these gaps mainly in following ways. First, developing theoretically justified normalization frameworks that clarify the information gain introduced by integrating source-side methods into target normalization. Second, incorporating dynamic field-level factors, such as publication growth rates, to further optimize metric performance. Third, exploring practical applications of the combined metrics, including identification of high-impact papers and journal evaluation, to assess their utility in real-world research assessment.

## Acknowledgments

The paper is a substantially extended version of the ISSI2025 conference paper (Lu et al., 2025). We thank Yahui Liu for valuable discussion.

## Competing interests

The authors declare no competing interests.

## Author contributions

**Xinyue Lu:** Data curation, Investigation, Visualization, Writing – original draft, Writing – review & editing.
**Li Li:** Data curation, Writing – review & editing.
**Zhesi Shen:** Conceptualization, Investigation, Visualization, Supervision, Writing – original draft, Writing – review & editing.

## Statement

During the preparation of this work the author(s) used LLM in order to improve readability and language. After using this tool/service, the author(s) reviewed and edited the content as needed and take(s) full responsibility for the content of the publication.